\pdfoutput=1
\documentclass[preprintnumbers,amsmath,amssymb,prd, notitlepage,nofootinbib,twocolumn,superscriptaddress]{revtex4-1}

\usepackage{graphicx}
\usepackage{dcolumn}
\usepackage{bm}
\usepackage{subfigure}

\usepackage{wasysym}
\usepackage{verbatim}
\usepackage{color}
\usepackage{todonotes}
\usepackage{float}
\usepackage{aasmacros}
\usepackage{placeins}
\usepackage[normalem]{ulem}

\presetkeys{todonotes}{inline}{}
\usepackage[italicdiff]{physics}

\newcommand{\st}{\sigma_\mathrm{T}}
\newcommand{\me}{m_\mathrm{e}}
\newcommand{\nume}{n_\mathrm{e}}
\newcommand{\rmn}{\mathrm}
\newcommand{\N}{\mathcal{N}}

\newcommand{\Mnu}{\Sigma m_{\nu}}
\newcommand{\tht}{\bm{\theta}}

\newcommand{\camb}{CAMB}

\newcommand{\tnp}{\top}
\newcommand{\lcdm}{$\Lambda$CDM}
\newcommand{\code}[1]{\texttt{#1}}

\begin{document}

\title{Improving Constraints on Fundamental Physics Parameters with the Clustering of Sunyaev-Zeldovich Selected Galaxy Clusters}

\author{Dylan Cromer}
\affiliation{Department of Astronomy, Cornell University, Ithaca, NY 14853, USA}
\author{Nicholas Battaglia}
\affiliation{Department of Astronomy, Cornell University, Ithaca, NY 14853, USA}

\author{Mathew S. Madhavacheril}
\affiliation{Department of Astrophysical Sciences, Princeton University, Princeton, NJ 08544, USA}

\date{\today}

\begin{abstract}

	Upcoming millimeter experiments that probe the cosmic microwave background (CMB) will observe tens of thousands of galaxy clusters through the thermal Sunyaev-Zeldovich (tSZ) effect. tSZ selected clusters are powerful probes of cosmological models, as they trace the late-time growth of structure. Late-time structure growth is highly sensitive to extensions to the standard cosmological model (\lcdm{}), such as the sum of the neutrino masses, the dark energy equation of state, and modifications to general relativity. The nominal statistic used for cluster observations is their abundances as a function of redshift. We investigate what additional cosmological information is gained after including the clustering signal of clusters, the cluster power spectrum. We forecast the cluster power spectra for the upcoming Simons Observatory and a CMB Stage-4-like experiment and find that the cluster power spectrum reduces marginalized constraints on the dark energy equation of state by $2$--$5\%$ and the growth index by around $2\%$, for example. We present the constraints using a generalized figure of merit and find improvements ranging from $4$--$7\%$ for extensions, $4$--$7\%$ for the astrophysical nuisance parameters, and $5$--$9\%$ for \lcdm{} parameters. We also find that if the bias of clusters as a tracer of the matter density can be measured to within $3\%$ or better, these improvements can be increased by up to a factor of $10$. We discuss the possibility of utilizing the clustering signal to address specific systematic uncertainties present in cluster abundance measurements.
	
\end{abstract}

\maketitle

\section{Introduction}
Galaxy clusters identified with the thermal Sunyaev-Zeldovich (tSZ) effect \citep{sz} provide a promising method for obtaining new constraints on the \lcdm{} model of cosmology and its extensions. The tSZ effect arises when cosmic microwave background (CMB) photons inverse-Compton scatter off of the hot intracluster medium (ICM), imprinting a unique spectral distortion in the CMB. These spectral distortions can be used to select locations of galaxy clusters on the sky, and follow up measurements with other wavelengths can then determine cluster redshifts. Galaxy clusters are excellent probes of late-time growth of structure, and so they are sensitive to parameters which influence structure growth, including extensions to the current \lcdm{} model such as the sum of neutrino masses, modifications to general relativity (GR) and the evolution of dark energy. This sensitivity has been shown by analysis of tSZ cluster catalogs \citep[e.g.,][]{Vik2009,spt-sz2010,Rapetti2010,act-sz2011,spt-sz2013, spt-sz2013b, act-sz2013, Mantz2014, Mantz2015,planck-sz2016,spt-sz2016,Bocquet2018}, and with simulations \citep[e.g.][]{matteo2016,matteo2018}.

Thermal SZ-selected clusters are advantageous due to the tSZ effect being nearly independent of redshift for clusters of fixed mass, and the selection function is easily modeled. While past tSZ cluster catalogs have been $\sim 10^2$ clusters \citep{Bleem2015,PlanckClust2016,Hilton2018}, stage 3 CMB experiments such as Advanced ACT \citep{advact}, SPT-3G \citep{SPT3G}, and the Simons Observatory (SO) \citep{simonsproj, simonsforecasts} will measure $\sim 10^4$ clusters, and the CMB Stage 4 (CMB-S4) experiment \citep{cmbs4} and potential future CMB satellites like PICO will measure $\sim 10^5$. These vastly expanded tSZ cluster catalogs will enable significant new constraints on \lcdm{} and its extensions \citep{LA2017,mat2017}. Overlap with optical surveys like the Large Synoptic Survey Telescope \citep[LSST][]{LSST} will allow for cluster confirmation and redshift determination through the red sequence \citep[e.g.][]{Gladders2001}.

A standard statistic for clusters is their abundances as a function of redshift, but as the sizes of tSZ cluster catalogs grow, their clustering statistics become useful probes. The two-point correlation function of clusters, the clustering power spectrum, gives information about the spatial correlations of clusters and cosmological parameters \citep{HH2003}. While limited by $\sqrt{N}$ sample variance in older tSZ cluster catalogs, for future CMB experiments the constraining power of the cluster power spectrum has been forecasted to have sensitivity to the sum of neutrino masses, the dark energy equation of state (EOS) and modifications to GR \citep[e.g.,][]{HH2003,clustering-mg2012} through the Kaiser effect \citep{kaisereffect}, and contains different cosmological information than cluster abundances. 

In this paper we forecast cosmological constraints obtained using both cluster abundances and the cluster power spectrum, with two models for the experimental sensitivities of SO \citep[representing their baseline and goal sensitivities and noise levels][]{simonsforecasts}, and one for a mock CMB-S4-like experiment \citep[using the noise levels in][]{mat2017}. These forecasts include a model for cluster selection used in \citet{mat2017}. We calculate the clustering Fisher matrix following the forecast methodology from \citep{fishcalc1997}, which has been used for the \emph{Euclid} satellite \citep{fishcalc2016} and future X-ray surveys \citep{fishcalc2010}. We treat the power spectrum as statistically independent from the cluster abundances, as we only use the large-scale, linear regime clustering information. We include the effects of photometric redshift errors through a Fourier-space kernel, and use a conservative redshift error estimate of $\sigma_z = 0.01$. We treat the cluster bias in two ways: first, we provide a forecast with the bias fixed at a known value, representing the maximal information contained in the power spectrum; second, we provide a forecast with the bias treated as a free parameter which is varied in the Fisher formalism. These are combined with cluster abundance forecasts from \citet{mat2017} as well as forecasts of the Planck primary CMB anistropies. Cosmological information is gained through sensitivity in the power spectrum to parameters; we do not use the constraints from clustering to calibrate cluster masses that are used in the cluster abundance forecasts \citep[e.g.,][]{Lima2004,majumdar2004}.

We forecast constraints on \lcdm{} parameters, extension parameters, and nuisance parameters and we re-express constraints on the dark energy EOS as constraints on the growth index $\gamma$. We use a generalized figure of merit (FoM) defined by \citep{defom} to characterize the constraining power on \lcdm{} parameters, extension parameters, and nuisance parameters. Finally, we discuss the results, and argue for the potential of clustering to mitigate certain systematic uncertainties present in cluster mass calibration through its constraining power on the nuisance parameters.

\section{Methodology} \label{sec:methods}

\subsection{Cluster Selection}

We use an analytic model for the tSZ signal for galaxy clusters, the integrated Compton-$y$. The tSZ spectral distortion of the observed CMB temperature is a function of frequency and the Compton-$y$ parameter ($y$): 

\begin{equation}
\frac{\Delta T(\nu)}{T_\rmn{CMB}} = f_\nu y,
\label{eq:delt_tsz}
\end{equation}
here $f_\nu = x\,\rmn{coth}(x/2) - 4$, where $x = h\nu / (k T_\rmn{CMB})$, $h$ is the Planck constant, and $k$ is the Boltzmann constant. Note that we neglected relativistic corrections to the tSZ spectral function $f_\nu$ \citep[e.g.,][]{Nozawaetal2006,Chluba2012}. The amplitude of the tSZ spectral distortion is directly proportional to $y$ (see Equation~\ref{eq:delt_tsz}), which is defined as, 
\begin{equation}
	y = \frac{\st}{\me c^2} \int \nume kT_\rmn{e} \dd{l},
\label{eq:y}
\end{equation}
and is proportional to the integrated electron pressure along the line-of-sight, $\dd l$. Here $\nume$ is the electron number density, $T_\rmn{e}$ is the electron temperature and $c$, $\me$, and $\st$ are physical constants corresponding to the speed of light, electron mass, and Thompson cross-section, respectively. We use an empirically measured pressure profile from \citep{Arnd2010} projected onto $y(\tht)$, where $\tht$ is the 2D angular coordinate on the sky.

For a given cluster we calculate its tSZ signal and observed signal-to-noise using a matched filter technique that exploits the unique spectral distortion of the tSZ effect \citep{Herranz2002,JB2006}. For the matched filter, the millimeter sky, $ \mathcal{M}(\tht)$, is modeled as
\begin{equation}
\mathcal{M}_\nu(\tht) = Y f_\nu g(\tht) + N_\nu(\tht),
\end{equation}
where $Y$ is the amplitude of the tSZ signal for a given halo, $g(\tht)$ is the normalized projected $y$ profile, $g(\tht) = y(\tht) / Y$, and $N_\nu(\tht)$ is the noise when searching for a tSZ signal. The noise is a function of $\nu$ and includes instrumental noise, atmosphere, primary CMB, and other secondary sources. The matched filter used to extract $Y$ is designed to minimize the variance across a given set of frequency bands for an assumed $y(\tht)$ profile
\begin{equation}
	\hat{Y} = \int F_\nu(\tht)^T \mathcal{M}_\nu(\tht) \dd{\tht} .
\label{eq:mf}
\end{equation}
Here and henceforth we assume that $\hat{Y}$ is an unbiased estimate of $Y$, we sum over $\nu$, and $F_\nu(\tht)$ is an unbiased, real-space matched filter that minimizes the variance. The equivalent form in Fourier space is,
\begin{equation}
F_\nu(\ell) = \sigma_N^2 \left[{C}_{N,\nu \nu'}(\ell)\right]^{-1} f_{\nu'} \tilde{g}(\ell),
\end{equation}
where $\tilde{g}(\ell)$ is the Fourier transform of the normalized projected $y$ profile, $\sigma_N^2$ is the variance, and $C_{N,\nu \nu'}(\ell)$ is the covariance matrix of the noise power spectrum. Note that the Fourier transform of $F_\nu(\ell)$ is $F_\nu(\tht)$. The variance is defined as 
\begin{equation}
	\sigma_N^2 = 2 \pi \int |\tilde{g}(\ell)|^2 f_\nu^T \left[{C}_{N,\nu \nu'}(\ell)\right]^{-1} f_{\nu'}\, \ell \, \dd{\ell},
\label{eq:erry}
\end{equation}
and the noise covariance matrix is defined as,
\begin{eqnarray}
	{C}_{N,\nu \nu'} (\ell) &=& {C}_{\mathrm{CMB},\nu \nu'} (\ell) + {C}_{\mathrm{sec},\nu \nu'} (\ell) \nonumber \\
	&+& \left(\frac{N_\nu(\ell)}{B_{\nu}(\ell)^2 }\right)\delta_{\nu \nu'}.
\end{eqnarray}
The value $\sigma_N$ represents the estimated error on $Y$ given the properties of the noise covariance matrix. Details on the components of the noise covariance matrix can be found in \citep{mat2017}, which uses the functional forms and parameters for additional secondary anisotropies from \citep{Dunkley2013}. The experimental properties used in our calculations such as frequency bands, beams, and noise for Simons Observatory and CMB-S4 are found in \citep{simonsforecasts} and \citep{mat2017}, respectively.

\subsection{Fisher Formalism for Clustering}

We use the Fisher formalism \citep{fisher} to forecast constraints on cosmological parameters. For the clustering Fisher matrix calculation we follow \citep{fishcalc2016, fishcalc2010, fishcalc1997}:
\begin{align}
	F_{\alpha \beta} &= \sum_{i, j, m} \pdv{\ln{\bar P(\mu_m, k_j, z_i)}}{p_\alpha} \pdv{\ln{\bar P(\mu_m, k_j, z_i)}}{p_\beta}  \notag
	\\
	&\qquad \times \frac{V^{\mathrm{eff}}_{i, j, m} k_j^2 \, \Delta k \, \Delta \mu}{8\pi^2} \notag
	\\
	&= \sum_{i, j, m} \pdv{\bar P(\mu_m, k_j, z_i)}{p_\alpha} \pdv{\bar P(\mu_m, k_j, z_i)}{p_\beta} \notag
	\\
	&\qquad \times \qty(\frac{V^{\mathrm{eff}}_{i, j, m} k_j^2 \, \Delta k \, \Delta \mu}{8\pi^2 \bar P^2(\mu_m, k_j, z_i)}), \label{eq:fisher}
\end{align}
with $k_j$ the norm of the wavevector $\va*{k}$, $m$ indexing $\mu = \cos(\theta)$ with $\theta$ the angle between $\va*{k}$ and the line of sight, $j$ indexing the $k$ bins, $i$ indexing redshift bins, $V_\mathrm{eff}$ is the effective volume of the survey, and $\bar P$ is the redshift-averaged cluster power spectrum. The clusters will be confirmed via optical surveys that overlap with the CMB surveys. These optical surveys will also provide photometric redshifts with around 1\% errors using red sequence techniques \citep[e.g.,][]{Gladders2001,Redmapper}.

We follow the notation and calculations in \citep{fishcalc2016}, with the exception that we use
\begin{equation}
	V^{\mathrm{eff}}(\mu_m, k_j, z_i) = V_0(z_i) \qty[ \frac{\tilde n(z_i) \bar P(\mu_m, k_j, z_i)}{1 + \tilde n(z_i) \bar P(\mu_m, k_j, z_i)}]^2,
	\label{eq:veff}
\end{equation}
which matches the formula in both \citep{fishcalc2010, fishcalc1997}. In Figure~\ref{fig:exampleplot} we show $\bar P$, the shot noise per mode $1/\tilde n$, and the total noise per mode $\bar P$ + $1/\tilde n$, which can be obtained from the factor in the parentheses of Equation~\ref{eq:fisher}. This factor is the inverse variance squared of the cluster power spectrum; the noise per mode is found by taking the reciprocal square-root of this factor and expanding $V^{\mathrm{eff}}$ with Equation~\ref{eq:veff}, then removing the prefactor which depends on the binning scheme used for $k$ and $\mu$. In Equation~\ref{eq:veff} $V_0(z_i)$ is the total comoving volume in the $i$th redshift bin, which we calculate via
\begin{equation}
	V_0(z_i) = 4 \pi \int_{z_i}^{z_{i+1}} \dd{z} \dv{V}{z}, \label{eq:v0}
\end{equation}	
and $\tilde n(z_i)$ is the average number density in this bin:
\begin{align}
	\tilde n(z_i) = \int_0^\infty &n(M, z_i) \N(\rmn{log}Y|\rmn{log}\bar{Y},\sigma_{\rmn{log}Y}) \notag
	\\
    &\times \frac{1}{2} \left[1 + \mathrm{erf}\left(\frac{Y - 5\sigma_N}{\sqrt{2}\sigma_N}\right) \right]  \dd{M} \dd{Y}. \label{eq:ntil}
\end{align}
Here $n(M, z_i)$ is the mass function using the functional form from \citep{Tink2008},
$\N(\rmn{log} Y|\rmn{log}\bar{Y},\sigma_{\rmn{log}Y})$ is a lognormal distribution of $Y$ given the mean integrated Compton-y ($\bar{Y}$), and the intrinsic scatter ($\sigma_{\rmn{log}Y}$), the selection function is defined by the standard error function ($\mathrm{erf}$) with a signal-to-noise threshold 5 using the estimated error, $\sigma_N$, from the matched filter (see Equation~\ref{eq:erry}).

The function for $\bar{Y}$ follows the scaling relation defined in \citet{mat2017},
\begin{align}
\bar Y(M,z) &= Y_\star \qty[(1-b) \frac{M_{500\rho_c}}{M_\star}]^{\alpha_Y} \exp[\beta_Y \log^2\qty(\frac{M_{500\rho_c}}{M_\star})] \notag
\\
&\times E^{2/3}(z) (1+z)^{\gamma_Y} \qty[\frac{D_A(z)}{100\rmn{Mpc}/h}]^{-2},
\end{align}
where $Y_\star = 2.42 \times 10^{-10}$, $M_\star = 10^{14} M_{\odot}/h$ is the pivot mass, $1-b$ parameterizes the mass bias, $\alpha_Y$ and $\beta_Y$ are the first and second order mass-dependence power laws, and $\gamma_Y$ parameterizes additional redshift dependence. $E(z) = H(z)/H_0$ is the Hubble function, and $D_A(z)$ is the angular diameter distance.

The function $\sigma_{\rmn{log}Y}$ follows the log-normal scatter model defined in \citet{mat2017}, 
\begin{equation}
\sigma_{\log{Y}}(M,z) = \sigma_{\log{Y},0} \qty[\frac{M_{500\rho_c}}{M_\star}]^{\alpha_\sigma} (1+z)^{\gamma_\sigma},
\end{equation}
with $\sigma_{\log{Y},0}$ the fiducial scatter, $\alpha_\sigma$ parameterizing the mass power-law, and $\gamma_\sigma$ the redshift dependence. Fiducial values and steps used in the Fisher analysis are shown in Table \ref{tab:params}.

The cluster power spectrum is also averaged between redshift bins:
\begin{equation}
	\bar P(\mu_m, k_j, z_i) = \frac{1}{S_i} \int_{z_i}^{z_{i+1}} \dd{z} \dv{V}{z} \tilde n^2(z) \tilde P(\mu_m, k_j, z), \label{eq:pbar}
\end{equation}
with $S_{i}$ and  $\tilde P$ again the same as in \citep{fishcalc2016},
\begin{align}
	S_i &= \int_{z_i}^{z_{i+1}} \dd{z} \dv{V}{z} \tilde n^2(z), \label{eq:sfunc}
	 \\
	 \tilde P(\mu_m, k_j, z_i) &= W^2(k_r) \qty[b_{\mathrm{eff}}(z_i) + f(z_i) \mu_m^2]^2 P_L(k_j, z_i), \label{eq:ptil}
\end{align}
where $b_{\mathrm{eff}}$ is the linear bias from \citet{Tink2010} weighted by Equation~\ref{eq:ntil} as defined in \citet{fishcalc2016}, $f$ is the growth rate $\eval{\dv{\ln D(a)}{\ln(a)}}_{a = (1+z)^{-1}}$ with $D(a)$ the linear growth factor, $W(k_r)$ is a radial Gaussian kernel accounting for photometric redshift errors, and $P_L$ is the linear power spectrum calculated here by \camb{} \citep{Lewis:1999bs,Howlett:2012mh} for the given cosmological parameters. 

We estimate photo-$z$ errors in the same manner as \citet{smith_etal_2018}, with the kernel $W(k_r)$ given by
\begin{equation}
W(k_r) = \exp(-\frac{\sigma_z^2}{2H(z_i)^2} k_r^2),
\end{equation}
where $\sigma_z^2$ is the redshift variance, $k_r = k_j \sqrt{1-\mu_m^2}$ is the radial magnitude of the wavevector $\va*{k}$, and $H(z_i)$ is the Hubble parameter at redshift bin $i$. This factor has the effect of suppressing power at large $k$, in particular where $k > H(z)/\sigma_z$. This corresponds to the distance scale of the redshift errors. The underlying cluster distribution is distorted by a single factor of $W$, so the power spectrum is distorted by $W^2$. Here, we use $\sigma_z = 0.01$ as a conservative estimate of available redshift precision for SO and CMB-S4 clusters. Note that we fix $H(z)$ to be at the fiducial cosmology when calculating derivatives.

We note that every quantity is computed strictly on a $\mu$, $k$, and $z$ grid. For computing Fisher matrices we use $9$ $\mu$ bins, with $\mu$ ranging from $-1$ to $1$; we use $133$ $k$ bins, ranging from $10^{-4} \, h/\mathrm{Mpc}$ to $0.14 \, h/\mathrm{Mpc}$; lastly for redshift we use $18$ $z$ bins, ranging from $0.1$ to $1.9$. Whereas \citet{majumdar2004} uses $k$s as high as $1.4 \, h/\mathrm{Mpc}$, we use a conservative range for $k$ so we are not sensitive to non-linear biasing, Fingers of God, non-linear matter power and baryonic feedback. This restricts the power spectrum to the large-scale regime, where we expect minimal covariance with cluster abundances \citep{masahiro2013,masahiro2014,Schaan2014}. To calculate the function values for the integrals between bins in Equations~\ref{eq:v0}--\ref{eq:sfunc} we chose to interpolate. These interpolations then allow the evaluation on finer grids to obtain sub-grid values.

There are uncertainties in the effective bias amplitude $b_{\mathrm{eff}}$. We address their effect on our constraints with two approaches: one where $b_{\mathrm{eff}}$ is determined exactly by the fitting function in \citet{Tink2010} and weighted by Equation~\ref{eq:ntil} to demonstrate the maximum amount of cosmological information, and a conservative one where we introduce a free parameter $a_{\mathrm{bias}}$ to scale the effective bias, and marginalize over this parameter. In this second approach, the cluster power spectrum takes the form
\begin{equation}
	\tilde P(\mu, k, z) = W^2(k_r) \qty[a_{\mathrm{bias}}b_{\mathrm{eff}}(z) + f(z) \mu^2]^2 P_L(k,z). \label{eq:ps_w_abias}
\end{equation}
In Equations \ref{eq:ptil} and \ref{eq:ps_w_abias}, the non-monopole ($\mu \neq 0$) spectra depend on the growth rate $f$ through the Kaiser effect \citep{kaisereffect}. The growth rate can be approximated as a function of the matter density $\Omega_m$ and the growth index $\gamma$ \citep{linder2005}, which governs how $f$ depends on $\Omega_m$ through the approximate equation
\begin{equation}
f \approx \Omega_m^{\gamma}.
\end{equation}
We include constraints on the dark energy equation of state parameter $w$ via a parameterized post-Friedmann evolution module included in CAMB \citep{ppf2007, ppf2008}. Deviations from $w=-1$ are parameterized as $w(a) = w + (1-a) w_a$, where we vary both $w$ and $w_a$ in the Fisher analysis. Constraints on $w$, $w_a$ can be used to constrain $\gamma$ using a fitting formula from \citep{linder2005}; we include the details of this procedure in Appendix \ref{appendix:growthindex}. Note that the constraints on $\gamma$ consider only cosmologies with standard gravity and evolving dark energy; we do not consider modified gravity theories when varying cosmologies.

\subsection{Cluster Abundances and Mass Calibration}  \label{subsec:abund}

We use forecasted cluster abundance Fisher matrices calculated by \citet{mat2017} as baseline constraints, to which we add clustering matrices. The cluster abundances per bin of Compton-$Y$ signal-to-noise ratio (SNR) $q_\rmn{Y}$, weak-lensing calibrated mass estimate $M_\rmn{L}$, and redshift, are modeled by
\begin{eqnarray}
\frac{N (M_\rmn{L},q_\rmn{Y},z)}{\Delta M_\rmn{L} \, \Delta q_\rmn{Y} \, \Delta z} &=& \int \frac{\dd^2 N}{\dd z \dd M} P(M_\rmn{L},q_\rmn{Y} | Y , M) \nonumber \\
& & \N(\rmn{log} Y|\rmn{log}\bar{Y},\sigma_{\rmn{log}Y}  ) \dd M \dd Y.
\label{eq:abund}
\end{eqnarray}
Here $P(M_\rmn{L},q_\rmn{Y} | Y , M)$ models the probability distribution $M_\rmn{L}$ and $q_\rmn{Y}$ given $Y$ and $M$, $\frac{\dd^2N}{\dd M\dd z}$ is the differential number of clusters with respect to $M$ and $z$, and $\N(\rmn{log} Y|\rmn{log}\bar{Y},\sigma_{\rmn{log}Y})$ is a lognormal distribution of $Y$ given $\bar{Y}$ and $\sigma_{\rmn{log}Y}$.

The model for $P(M_\rmn{L},q_\rmn{Y} | Y , M)$ is two independent normal distributions ($\N$), 
\begin{eqnarray}
\label{eq:pq}
P(M_\rmn{L},q_\rmn{Y} | Y , M) &=& \N (q_\rmn{Y} | Y/\sigma_N , 1)  \nonumber \\
&\,& \N(M_\rmn{L}|M,\sigma_M).
\end{eqnarray}
The estimated $Y$ error, $\sigma_N$ is determined from the matched filter (see Equation~\ref{eq:erry}). The calibration masses $M_\rmn{L}$ are obtained via optical weak-lensing signals modeled by calculating the excess surface density projected along the line of sight, assuming a Navarro-Frenk-White \citep[NFW,][]{nfw1997} density profile computed from the cluster mass $M$. The details of this procedure are elaborated further in \citep{mat2017}.

\section{Results} \label{sec:results}

We present forecasts for the Simons Observatory baseline and goal noise levels and forecasts for a CMB-S4-like experiment. We illustrate the constraints from clustering by comparing Fisher forecasts of the Planck primary CMB anisotropies added to SO/S4 tSZ cluster abundances Fisher matrices from \citep{mat2017}\footnote{The tSZ cluster abundances Fisher matrices that we use in this work differ slightly from the results in \citep{mat2017} for $w$ and $w_a$, which is the result of including PPF formalism}. The cluster abundances Fisher matrices are derived from a Poisson likelihood and the clusters are binned as a function of signal-to-noise, weak-lensing mass calibration, and redshift. We include a prior on the optical depth to reionization, $\tau$, of $\sigma(\tau) = 10^{-2}$. This added prior accounts for additional low-$\ell$ polarization power spectrum constraints on $\tau$ obtained by Planck, which are not included in the Planck forecast.

\begin{table}[!htbp]
\begin{tabular}{llll}
	\hline
	Parameter         & Fiducial            & Step                 & Prior      \\
	\hline
	$H_0$             & $67.0$              & $0.5$                &            \\
	$\Omega_bh^2$     & $0.022$             & $0.0008$             &            \\
	$\Omega_ch^2$     & $0.1194$            & $0.003$              &            \\
	$A_s$             & $2.2\times 10^{-9}$ & $0.1 \times 10^{-9}$ &            \\
	$n_s$             & $0.96$              & $0.01$               &            \\
	$\tau$            & $0.06$              & (Not varied)         & $10^{-2}$  \\
	\hline
	$\Sigma m_{\nu}$  & $0.06$              & $0.02$               &            \\
	$w$               & $-1$                & $0.01$               &            \\
	$w_a$             & $0.0$               & $0.05$               &            \\
	\hline
	$b$               & $0.8$               & $0.1$                &            \\
	$\alpha_y$        & $1.79$              & $0.04$               &            \\
	$\sigma_{\log Y}$ & $0.127$             & $0.02$               &            \\
	$\gamma_{Y}$      & $0$                 & $0.02$               &            \\
	$\beta_Y$         & $0$                 & $0.02$               &            \\
	$\gamma_\sigma$   & $0$                 & $0.02$               &            \\
	$\beta_\sigma$    & $0$                 & $0.02$               &            \\
	\hline
	$a_\mathrm{bias}$ & $1$                 & $0.01$               &            \\
	$\sigma_z$        & $0.01$              & (Not varied)         &
	          \\
	\hline
\end{tabular}%
\caption{\label{tab:params} Fiducial parameters and their step values used in calculating the Fisher matrix. The step sizes used are the same as those used by \citet{mat2017} and are tested for numerical stability. The top section shows \lcdm{} parameters, the second section shows extensions, the third astrophysical nuisance parameters, and at the bottom is the cluster bias variance parameter and redshift error level.}
\end{table}

\begin{figure}[!htbp]
        \includegraphics[width=\linewidth,height=\textheight,keepaspectratio]{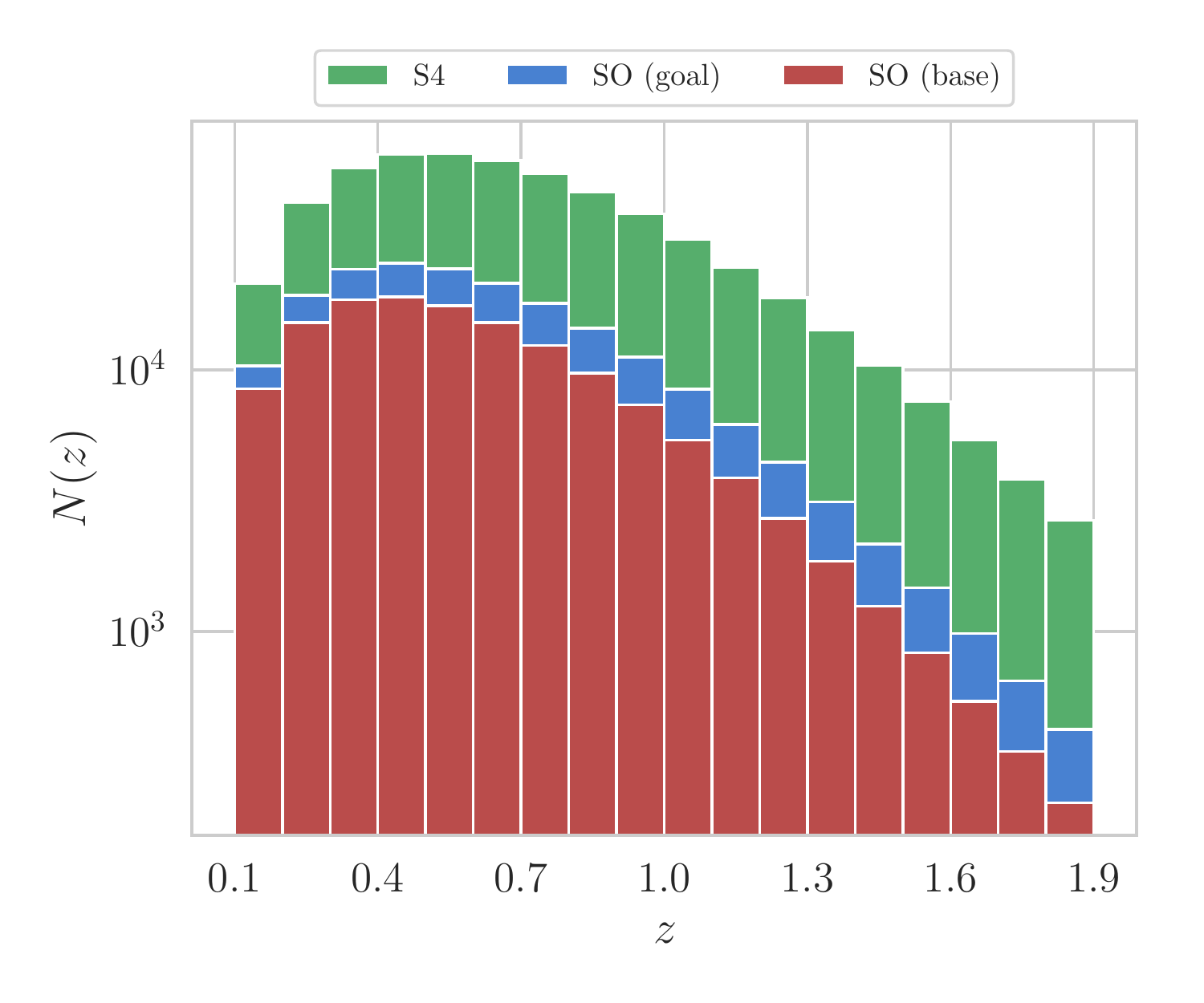}
	\caption{\label{fig:num_clusters}
    Abundance of clusters as a function of redshift for each experiment. The abundance $N(z)$ is the forecasted number of clusters in each redshift bin selected by the given experiment, calculated using the method described in Section \ref{subsec:abund}.
	}
\end{figure}

\begin{figure}[!htbp]
        \includegraphics[width=\linewidth,height=\textheight,keepaspectratio]{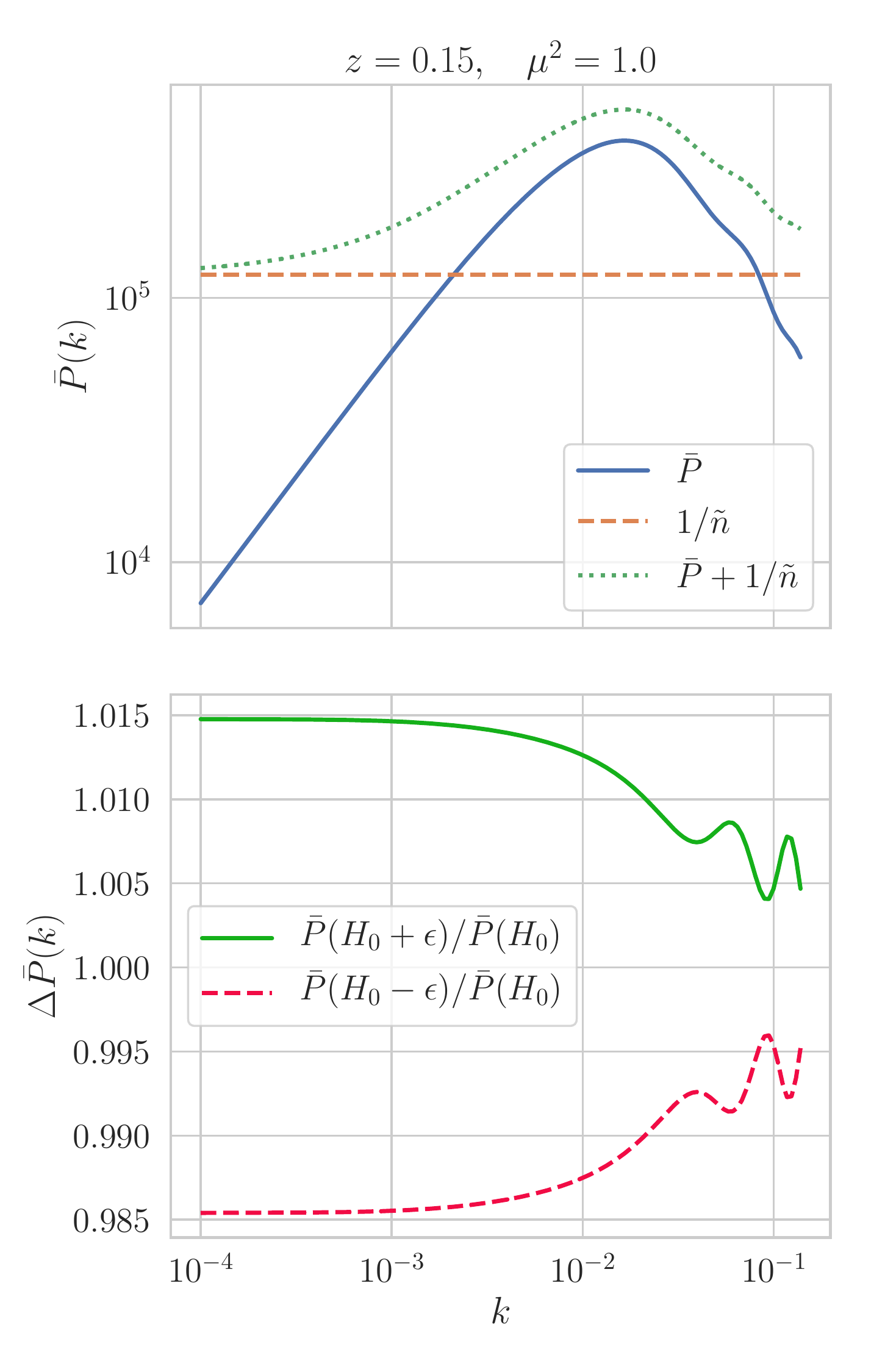}
	\caption{\label{fig:exampleplot}
		Top: The cluster power spectrum, $\bar P(z=0.15, \mu=1)$ (0.15 is the center of the first redshift bin), is plotted with the shot-noise per mode $1/\tilde n$, and the total noise per mode $\bar P + 1/\tilde n$. Bottom: The differences between spectra at the fiducial value of $H_0=67.0$ and the $H_0 \pm \epsilon$ values. Shifts in $\bar P$ are due to changes in $H_0$ shifting the scale of the baryon acoustic oscillation modes and the sound horizon.
	}
\end{figure}

\begin{figure}[!htbp]
        \includegraphics[width=\linewidth,height=\textheight,keepaspectratio]{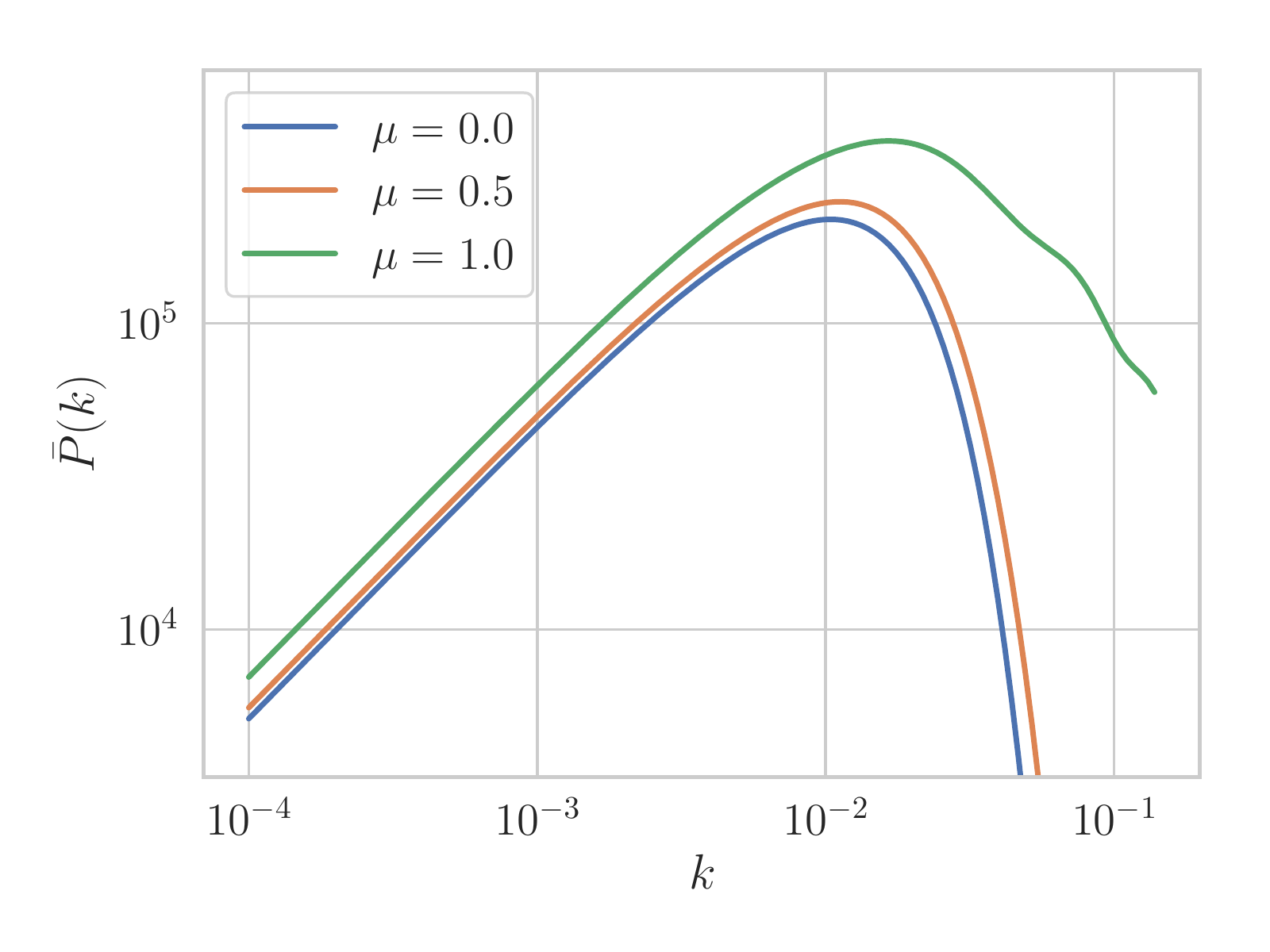}
	\caption{\label{fig:photoz}
		Cluster power spectrum for $z=0.15$ (the center of the first redshift bin) at various values of $\mu$. The primary effect shown is the effect of photometric redshift errors (here we use $\sigma_z = 0.01$); the $\mu = 1$ (perpendicular) spectrum is completely unaffected by the photo-$z$ errors, and the $\mu = 0$ (radial) spectrum is affected the most by this source of error. Damping at $k$ greater than $H(z)/\sigma_z$ is visible for the $\mu = 0,0.5$ spectra. The effect of changing $\mu$ also alters the overall amplitude due to the contribution of RSD in Equation \ref{eq:ptil}.
	}
\end{figure}

\begin{figure*}[!tbp]
        \includegraphics[width=\textwidth,height=\textheight,keepaspectratio]{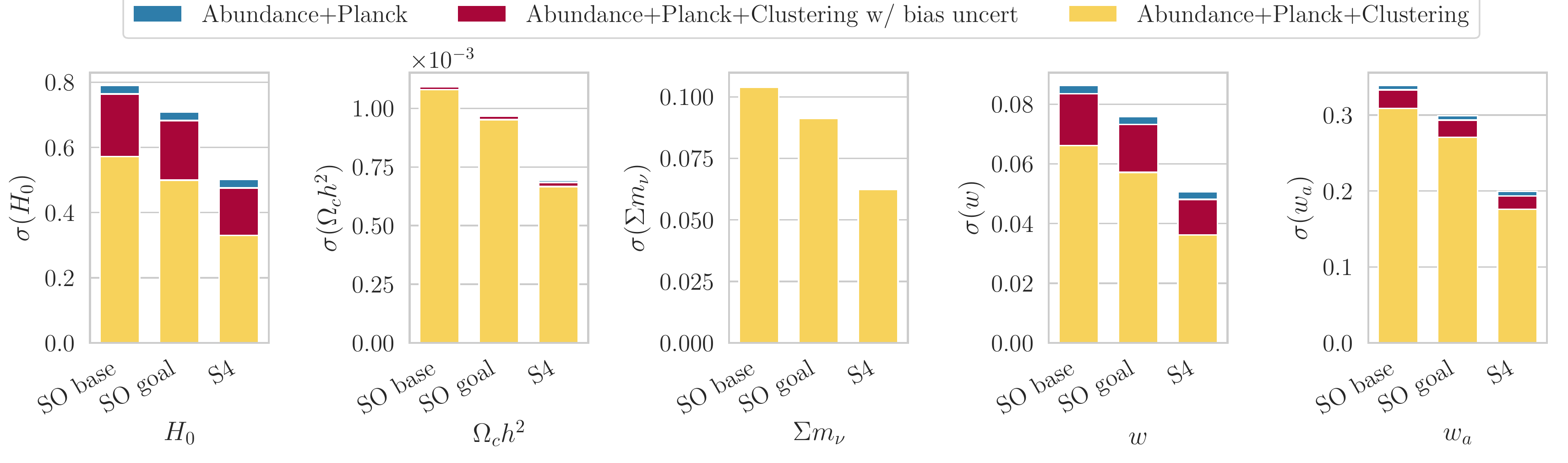}
	\caption{\label{fig:barplot}Total marginalized errors for a selection of parameters, illustrating the effects of adding clustering information. The blue bars show the errors found when including primary CMB information from Planck combined with tSZ cluster abundances; the red and yellow bars show the conservative, and optimal cases respectively, of adding information from clustering. In some cases the conservative case of allowing variance in the cluster bias has a large impact on the error (e.g. for $w$), but in others it does not (e.g. neutrino masses).}
\end{figure*}

\begin{figure}[!htbp]
	\includegraphics[width=0.70\linewidth,height=0.70\textheight,keepaspectratio]{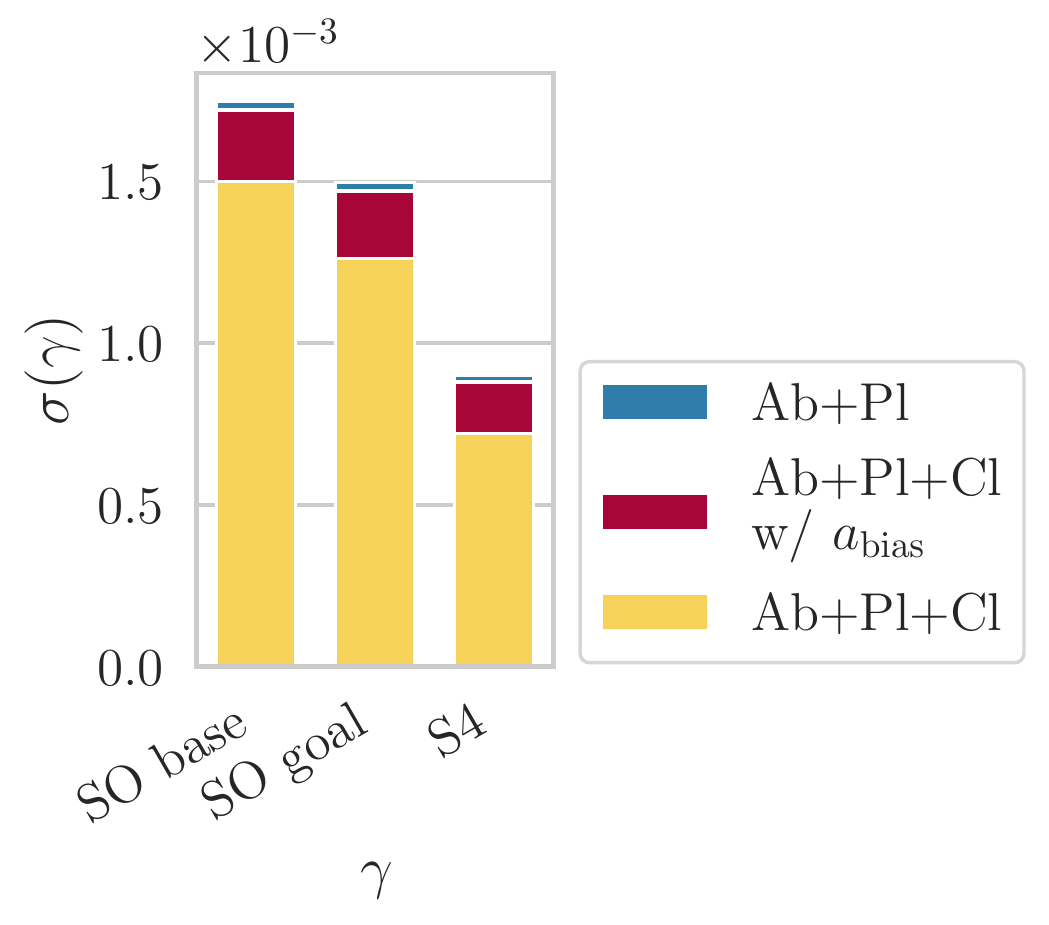}
	\caption{\label{fig:gammabarplot}
		Marginalized error for $\gamma$ obtained via the Jacobian discussed in Section~\ref{sec:methods} and Appendix~\ref{appendix:growthindex}, for the same three cases as in Figure ~\ref{fig:barplot}. Note that these constraints are derived from the constraints on $w$ and $w_a$ from the fitting formula in Equation~\ref{eq:gamma_fit_formula}.
	        }
\end{figure}

We illustrate the redshift distribution of clusters in Figure \ref{fig:num_clusters}, where we plot total cluster abundances as a function of redshift, as well as post-SZ-selection cluster abundances for each experiment.

We show an example of the differences in power spectra when shifting the Hubble parameter in the bottom panel of Figure~\ref{fig:exampleplot}. The differences between Hubble parameter values is the same step size we used for computing numerical derivatives. These differences between $\bar P$ are due to shifting the scale of the baryon acoustic oscillation modes and the scale of the sound horizon as we change $H_0$. The symmetry between the spectrum evaluated at $H_0 + \epsilon$ and $H_0 - \epsilon$ can be understood by Taylor-expanding $\bar P$ about $H_0$, giving the difference to first order as $\delta \bar P \approx \epsilon \partial_{H_0} \bar P$. Thus changing the sign of $\epsilon$ will produce a nearly symmetrical shift in $\bar P$ for small step sizes. In Table \ref{tab:params} we show the full list of parameters used, along with their step sizes.

We also examine the impact of photometric redshift errors on the power spectrum. In Figure \ref{fig:photoz}, we show the spectrum at three values of $\mu$. Because the photo-$z$ errors only change the radial component of the power spectrum, for $\mu = 1$ there is no contribution from these errors. At $\mu = 0$ the effect is maximized, and we see damping at large $k$. This reduces the available cosmological shape information, as much of the shape variation in $\bar P$ occurs at these large $k$ values. It does not reduce amplitude information, as changes in the amplitude are due to RSD and not photo-$z$ errors.

We characterize improvement in the marginalized errors on parameters using
\begin{equation}
100 \times \qty(1 - \frac{\sigma(\mathrm{Ab+Pl+Cl})}{\sigma(\mathrm{Ab+Pl})}),
\end{equation}
with $\sigma(\mathrm{Ab+Pl+Cl})$ the marginalized error from clustering, abundances, and Planck primary CMB, and $\sigma(\mathrm{Ab+Pl})$ the error from only abundances and Planck. The marginalized constraints are improved by the addition of clustering in all cases. Figure \ref{fig:barplot} illustrates the marginalized errors for selected parameters, including Fisher matrices for the Planck satellite and for cluster abundances (as obtained in \citep{mat2017}). Overall, $w$ and $H_0$ show the highest gains in information. We express these constraints as ranges with a lower bound of improvement given by the conservative case where the cluster bias is unknown, and the maximally informative case where it is known exactly. We find that for SO baseline noise levels, a $3$--$23\%$ range improvement for $w$, and a $2$--$9\%$ range for $w_a$. For SO goal noise levels these ranges are $4$--$25\%$ ($w$) and $2$--$9\%$ ($w_a$); for CMB-S4, $5$--$29\%$ ($w$) and $3$--$12\%$ ($w_a$). Clustering is less sensitive to the neutrino mass sum than to the dark energy parameters, with all improvements of $\sigma(\Mnu)$ roughly $0.5\%$ for all experiments. Furthermore, the improvements change very little between the conservative case and the maximal case for cluster bias, with differences in improvement of around $0.1\%$ or less.

The Hubble rate $H_0$ improves by $3$--$28\%$ for SO baseline, $4$--$30\%$ for SO goal, and $5$--$34\%$ for CMB-S4. The matter densities $\Omega_m h^2$ and $\Omega_b h^2$, are improved by $0.2$--$4\%$. The scale-dependence of the linear matter power spectrum, $n_s$, is only improved $\sim 03$--$1\%$. Lastly, $A_s$ is improved only by $0.2$--$0.8\%$.

Using the methods discussed in Section \ref{sec:methods} and in Appendix \ref{appendix:growthindex}, we convert the constraints on $w,w_a$ to constraints on $\gamma$. The marginalized error on $\gamma$ is shown in Figure \ref{fig:gammabarplot}, with improvements of $2$--$14\%$ (SO baseline), $2$--$16\%$ (SO goal), and $2$--$20\%$ (CMB-S4). While we cannot compute an improvement for $a_\mathrm{bias}$, we can report constraints in our conservative forecast, with absolute marginalized errors of $0.019, 0.017$, and $0.010$, corresponding to SO baseline, SO goal, and CMB-S4 respectively.

\begin{figure*}[!tbp]
        \includegraphics[width=\textwidth,height=\textheight,keepaspectratio]{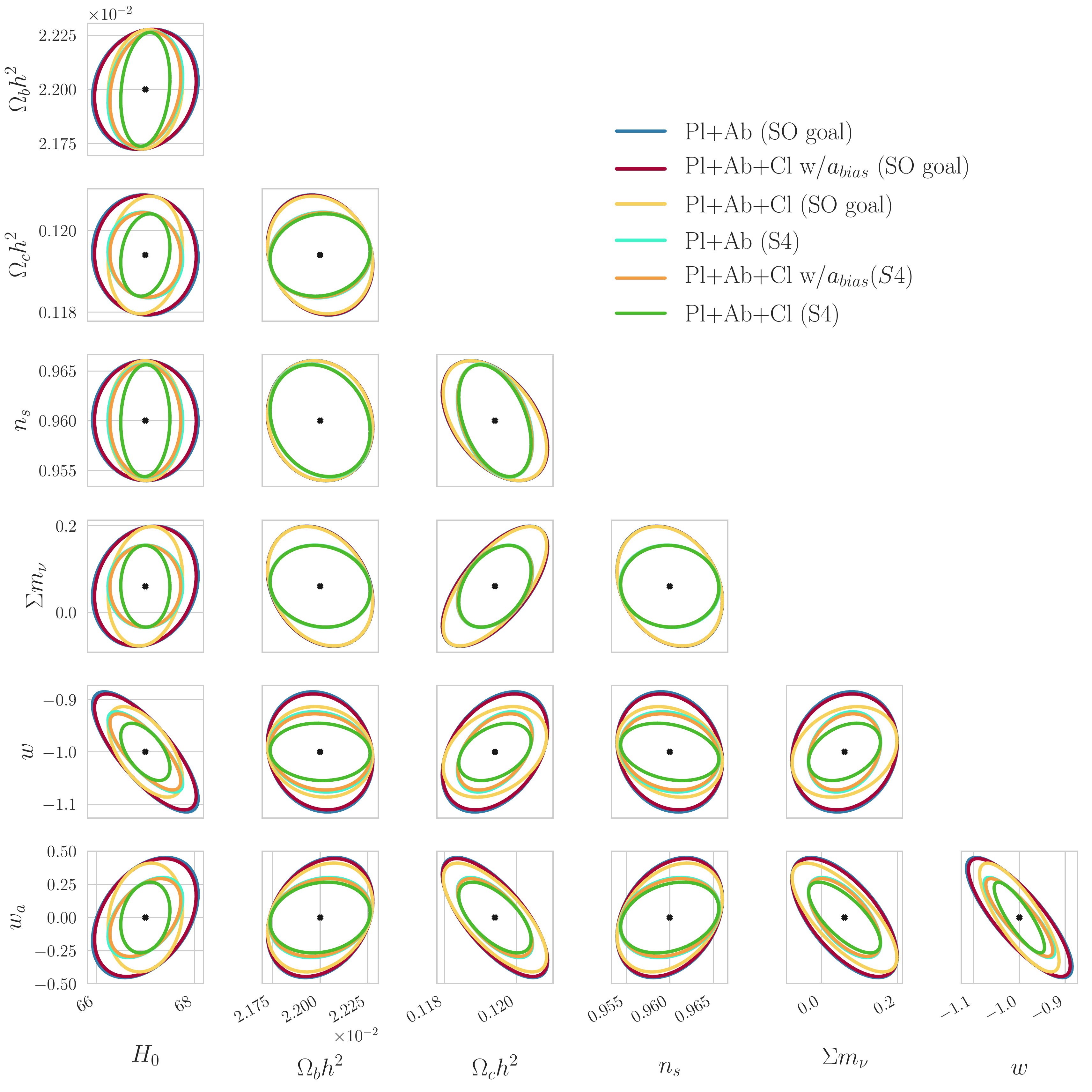}
	\caption{\label{fig:triplot} Ellipses showing the $68.3\%$ confidence regions for selected parameter pairs. In the legend, Pl+Ab refers to Planck primary CMB plus cluster Abundance constraints; Cl refers to adding clustering, and w/$a_\mathrm{bias}$ refers to the inclusion of uncertainty in the cluster bias. Note here only the goal noise-level SO and CMB-S4 experiments are plotted (SO baseline is omitted for figure clarity). One notable feature is that in general, the improvements in joint constraints are larger than the improvements in marginalized errors for many parameters.}
\end{figure*}

If we do not marginalize all parameters, we find better constraints on extensions. One way to show this is with joint-confidence ellipses, shown in in Figure \ref{fig:triplot}. A figure of merit (FoM), which in 2 dimensions is proportional to the inverse area of the confidence ellipses, helps indicate the amount of information gained on joint constraints for two parameters. For the $(w,w_a)$ ellipse, the FoM increases by $4$--$35\%$ for SO baseline, $4$--$39\%$ for SO goal, and $6$--$52\%$ for CMB-S4 with the addition of clustering (again with the range indicating from conservative knowledge of the cluster bias, to perfect knowledge). The $(\Mnu,w)$ ellipse shows improvements of $4$--$33\%$ (SO baseline), $4$--$36\%$ (SO goal), and $6$--$46\%$ (CMB-S4); the $(\Mnu,w_a)$ ellipse improves by $3$--$24\%$ (SO baseline), $4$--$26\%$ (SO goal), and $5$--$28\%$ (CMB-S4). This shows that clustering contains information on $w$, $w_a$, and $\Mnu$ that is not apparent in the fully marginalized constraints.

\begin{figure*}[!tbp]
        \includegraphics[width=0.8\textwidth,height=0.8\textheight,keepaspectratio]{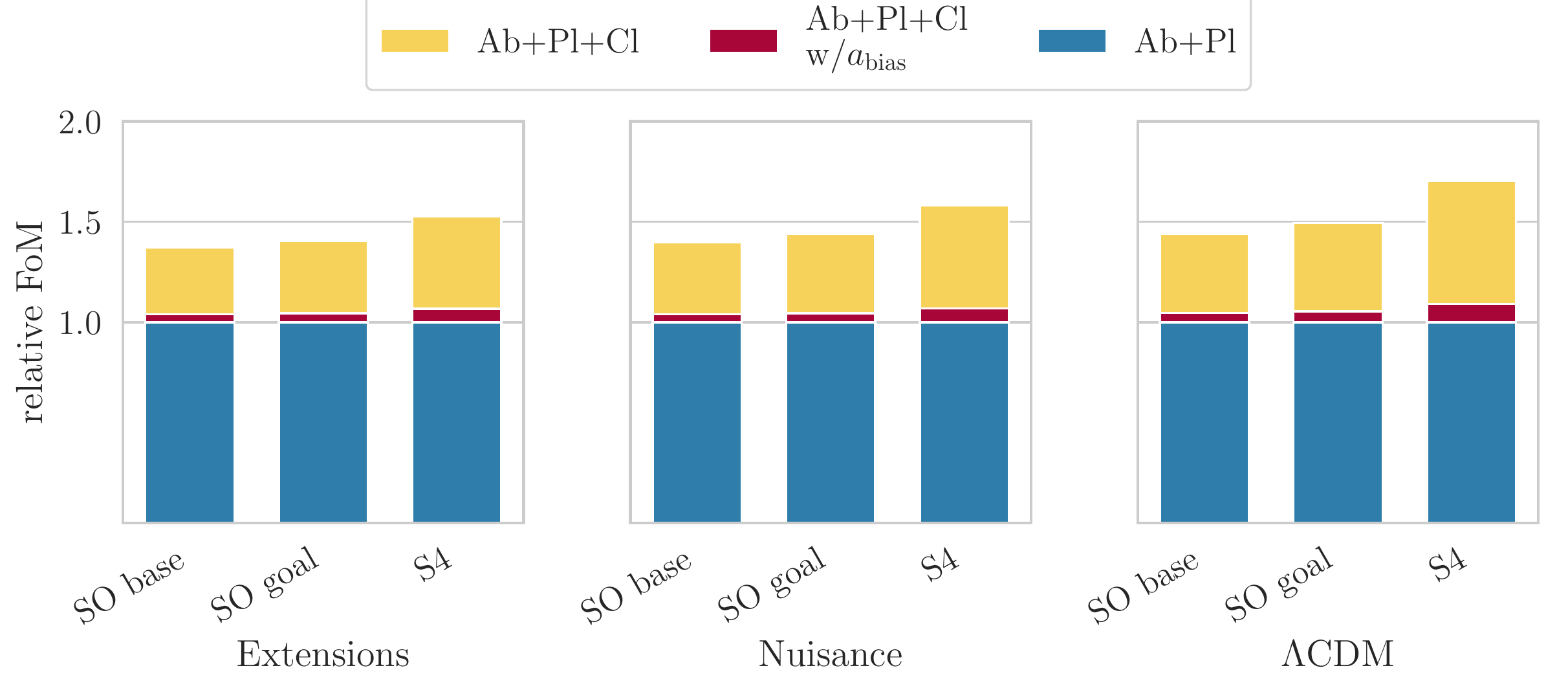}
	\caption{\label{fig:fom}Relative FoM for three sets of parameters, calculated for each experiment. Shown is the ratio of the FoM for the given forecast with the forecast for Planck primary CMB and abundances without clustering (hence the blue bars all show relative FoM of one). The red bars shows conservative gains from clustering, uncertainty in the cluster bias parameterized through $a_{\mathrm{bias}}$, while the yellow bars show the maximal gains if the bias is known exactly. The parameters included in each subfigure are: $\Mnu, w, w_a$ (extensions); $\alpha_y,\sigma_{\log Y},\gamma_{Y},\beta_Y,\gamma_\sigma, \beta_\sigma$ (nuisance), and $H_0, \Omega_c h^2, \Omega_b h^2, A_s, n_s$ (\lcdm{}).}
\end{figure*}

Going beyond pairs of parameters, in order to show the information gains from clustering for the broader sets of \lcdm{}, extension, and nuisance parameters without marginalizing, we utilize a generalized FoM defined in \citep{defom}:
\begin{equation}
	\mathrm{FoM}(p_1, p_2, \dots) = \frac{1}{\sqrt{\det{C(p_1, p_2, \dots)}}},
\end{equation}
where $p_1,\dots$ are parameters and $C(p_1, \dots)$ is the covariance matrix for that subset of parameters. The generalized FoM is proportional to the inverse (hyper) volume of the confidence region for the specified parameters. We calculate the FoM relative to the FoM for Planck primary CMB and cluster abundances alone, plotting the results in Figure \ref{fig:fom}. The improvements shown are, for extensions ($\Mnu, w, w_a$), $4$--$37\%$ (SO baseline), $5$--$41\%$ (SO goal), and $7$--$53\%$ (CMB-S4); for nuisance parameters ($\alpha_y,\sigma_{\log Y},\gamma_{Y},\beta_Y,\gamma_\sigma, \beta_\sigma$), $4$--$40\%$ (SO baseline), $5$--$44\%$ (SO goal), and $7$--$58\%$ (CMB-S4); and lastly, for \lcdm{} parameters ($H_0, \Omega_c h^2, \Omega_b h^2, A_s, n_s$), $5$--$44\%$ (SO baseline), $5$--$49\%$ (SO goal), and $9$--$71\%$ (CMB-S4).

\begin{figure*}[!tbp]
        \includegraphics[width=\textwidth,height=\textheight,keepaspectratio]{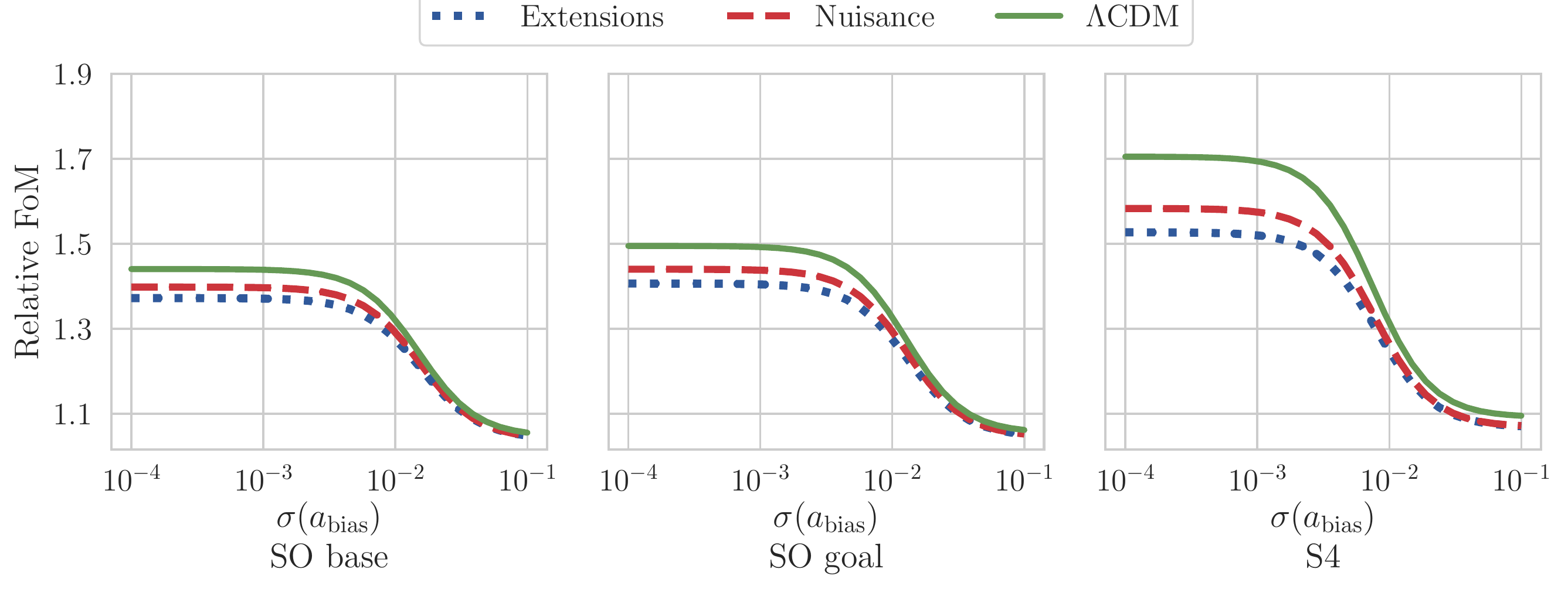}
	\caption{\label{fig:fom_of_abias} Relative FoM as a function of a prior on $a_\mathrm{bias}$. Extensions, nuisance parameters, and \lcdm{} parameters refer the same sets of parameters as in Figure \ref{fig:fom}. Once a prior of $\sigma(a_\mathrm{bias}) \sim 0.03$ or smaller is reached, the information added by clustering increases rapidly until $\sigma(a_\mathrm{bias})$ is roughly $0.003$, where this improvement begins to saturates.}
\end{figure*}

We address the possibility of attaining the maximal information from clustering, through an external measurement of the cluster bias. Here we add priors on $a_\mathrm{bias}$ and examine the relative FoM for each set of parameters parameters as a function of the prior $\sigma(a_\mathrm{bias})$, as shown in Figure \ref{fig:fom_of_abias}. We find that priors of $\sigma(a_\mathrm{bias}) \gtrsim  0.03$ do not add significant information. For $0.003 < \sigma(a_\mathrm{bias}) < 0.03$, the FoM values rapidly increase towards the values they take in the maximally informative case. Once $\sigma(a_\mathrm{bias}) \sim 0.003$, the figures of merit approach their asymptotic values, and improvements saturate.

We also examine the effects of $\tau$ on $\sigma(\Mnu)$ \citep{allison_etal_2015}, by calculating the constraints on $\Mnu$ as a function of the $\tau$ prior. In \citet{planck2016constraints}, the errors on $\tau$ are $\sigma(\tau) \sim 0.01$; we found that priors of $\sigma(\tau) \lesssim 0.1$ improve $\sigma(\Mnu)$, until about $\sigma(\tau) \sim 0.003$. We examine $\sigma(\Mnu)$ with all extensions, with only $\Mnu$ and $w$ present (removing $w_a$ from the Fisher matrix), and finally with only $\Mnu$, in order to examine whether degeneracies between $\Mnu$ and dark energy are concealing sensitivity to uncertainty in $\tau$. In all cases we find our results are consistent with those in \cite{allison_etal_2015}, and $\sigma(\tau) \sim 0.003$ is close to saturating $\sigma(\Mnu)$ around the cosmic variance limit.

\section{Discussion} \label{sec:discussion}

Current and future CMB experiments promise SZ-selected cluster catalogs of considerably larger size than those obtained from current-generation experiments. In particular, Advanced ACT \citep{advact}, SPT-3G \citep{SPT3G}, and the Simons Observatory (SO) \citep{simonsproj, simonsforecasts} will push the number of clusters to tens of thousands while CMB-S4 will push to hundreds of thousands. These increases promise improved constraints on \lcdm{} and its extensions. Overlap with the Large Synoptic Survey Telescope optical survey will allow confirmation of these catalogs along with redshift measurements.

We investigated the sensitivity of the clustering signal from galaxy clusters to cosmological parameters independent from their abundances. To do this, we performed a Fisher forecast of the cluster power spectrum. Our analysis is similar to \citet{majumdar2004}, however we have differed in several ways: We cut off the cluster power spectrum at $k = 0.14 \, h/\rmn{Mpc}$, an order of magnitude smaller than their maximum $k$; they did not include the Kaiser effect in their calculations; they utilized the amplitude of the power spectrum to infer the cluster bias as a function of mass, and used this indirect mass measurement as a calibrator for cluster abundances, whereas we only use the parameter sensitivity of the power spectrum; furthermore, the cluster abundances used here are inferred using the mass function from \citet{Tink2010}, as well as a sophisticated treatment of the SZ selection function from \citep{mat2017}.

In our analysis, we find constraints on extensions to \lcdm{}, including a dynamical dark energy equation of state and a nonzero neutrino mass sum, quantified by a relative figure of merit. These constraints improve with the addition of clustering by at least $\sim 4$--$5\%$ for the Simons Observatory (ranging over baseline and goal noise levels), with the potential for up to $\sim 41\%$ if the effective cluster bias can be calibrated using another measurement. For CMB Stage-4, we find improvements of at least $7 \%$ in the figure of merit for extensions, with a maximum of $53\%$. We also obtained constraints on \lcdm{} parameters, ranging from at least $5\%$ and up to $71\%$ across experiments. Constraints on nuisance parameters which control the cluster selection model are also improved by $4$--$58\%$.

It is apparent from the results shown that the statistical power of the clustering signal will depend strongly on how well the effective cluster bias variance can be constrained. Our idealized case with $a_{\mathrm{bias}}$ fixed to 1 includes information from both the shape and overall amplitude of the power spectrum, while when $a_{\mathrm{bias}}$ is a free parameter, the amplitude information is limited due to degeneracy with $b_\mathrm{eff}$ (see Equation \ref{eq:ps_w_abias}).

The Fisher analysis obtained constraints for $a_\mathrm{bias}$, with $\sigma(a_\mathrm{bias}) \sim 10^{-2}$. By adding priors for $\sigma(a_\mathrm{bias})$, we found that the error levels begin decreasing with a prior of $\sigma(a_\mathrm{bias}) \sim 0.03$, and continue until saturation at $\sigma(a_\mathrm{bias}) \sim 0.003$. Thus if the cluster bias can be measured to $3 \%$ or better through a different observable, clustering measurements can do better than our conservative case. Improvements saturate when the bias is measured to around $0.3\%$, so any measurement of the bias with precision $3$--$0.3\%$ will improve constraints from clustering. If the cluster bias is calibrated in this way, the constraining power added can be increased by up to a factor of 10. One possibility for external calibration of the bias is through cross-correlation with CMB lensing (e.g. \citep{PlanckBias}). 

We included the effects of photometric redshift errors on the predicted constraining power of the cluster power spectrum. For the main results, we used a conservative error level of $\sigma_z = 0.01$. For a more optimistic level of $\sigma_z = 0.005$, we found the percent improvement of the figure of merit improves by about $1$--$3\%$ in the conservative case with unconstrained bias, and $10$--$20\%$ when the bias is fully constrained. Where redshift errors have the strongest impact the results is at high $k$. At these $k$ the cosmological parameters tend to impact the shape of the power spectrum, which is damped away by redshift errors, in particular when we have no prior on the cluster bias. They less strongly affect the optimistic forecast where we saturate our knowledge of the $a_\mathrm{bias}$ parameter because we are able to recover more information from the overall amplitude and the Kaiser effect. This indicates that the power spectrum amplitude contains most of the information available for constraints, at least at the $1\%$ redshift error level.

Does the cluster power spectrum offer any advantages over the galaxy power spectrum, like one calculated from an LSST-like survey? Cluster catalogs are smaller and will have higher shot-noise. However, their redshift errors are considerably lower compared to photometric galaxy surveys. Here clusters are aided by the fact that they contain multiple photometric redshift estimates and the red sequence \citep[e.g.,][]{Gladders2001,Redmapper} compared to photometric redshift estimates for a single galaxy. Together these enable more precise and accurate redshift for clusters. A next step in understanding the improvement from joint abundance and clustering measurements is to calibrate cluster masses with the clustering measurement \citep{Lima2004,majumdar2004} through a linear bias relationship \citep[e.g.][]{Tink2010}. Additionally, this would require the clustering measurement to be made on the given galaxy cluster sample rather than a larger galaxy sample.

Another interesting aspect of our results is the additional information clustering provides on the nuisance parameters of the cluster selection model. The improvements in FoM are at least $4\%$ once clustering is added to SO, and at least $7\%$ once added to CMB-S4. Currently, constraints from tSZ cluster abundances are limited by systematics in the mass calibration and not by noise. The clustering signal may be able to reduce the impact of these systematics when combined with abundances. The cluster selection model already accounts for some systematics such as mass and redshift dependence in the Compton $Y$-cluster mass scatter. However, more work remains to evaluate whether clustering is a useful tool for reducing systematics in mass calibrations, which requires one to move beyond the Fisher formalism to a full, simulated likelihood function analysis.

A further way to extend this work would be to fully treat modified gravity when varying cosmologies; we estimate constraints on the growth index $\gamma$ from constraints on evolving dark energy, but to properly predict the sensitivity of the cluster power spectrum to modified gravity, it should be varied separately from the dark energy EoS parameters. One way to do this would be to allow the growth index to vary independently. We also did not include the Alcock-Paczynski effect \cite{ap1979} in our forecast. In principle this offers another way to extract information on the Hubble parameter and dark energy parameters from the power spectrum.

In summary, our results show that the clustering signal of tSZ-selected clusters can provide significant new constraining power in future CMB experiments such as the Simons Observatory and CMB Stage-4. This constraining power offers improvements for \lcdm{} parameters, extensions to \lcdm{}, and could potentially be used to reduce systematic uncertainties currently limiting the power of cluster abundances. Furthermore, the power of these results can be increased by calibrating clustering using a separate measurement of the effective linear bias of clusters accurate to at least $3\%$.

\begin{acknowledgments}

We thank Georgios Valogiannis for discussions regarding constraints to modified gravity. We thank Matteo Cataneo, Tom Crawford, Simon Foreman, Colin Hill, Eduardo Rozo, and Emmanuel Schaan for their comments on this work. 

This work is not an official Simons Observatory paper or CMB-S4 paper.

For the calculations in this work we use a public code library SZ Astrophysics Routines, \code{github.com/nbatta/szar}, an open source Python library used previously to forecast constraints from cluster abundances \citep{mat2017}.

\end{acknowledgments}

\bibliographystyle{apsrev}

\bibliography{dc}

\appendix

\section{Constraining the Growth Index} \label{appendix:growthindex}

To collapse our constraints on the dark energy EOS parameters $w,w_a$ to a constraint on the growth index $\gamma$, we use the fitting formula for $\gamma$ in \cite{linder2005}:

\begin{align}
        \gamma &= 0.55 + 0.05\qty[1 + w(z=1)]
        \\
        &= 0.55 + 0.05\qty[1 + w + \frac{w_a}{2}]. \label{eq:gamma_fit_formula}
\end{align}

The standard method for transforming variables in the Fisher formalism is via the Jacobian $J_{p,p'} = \pdv{p}{p'}$, with $p'$ the new parameters and $p$ the old ones \citep{coefisher}. In this case, we project two parameters $w$ and $w_0$ down to one, $\gamma$, so the Jacobian is non-square. Unlike a usual coordinate transformation this is non-invertible, as one value for $\gamma$ corresponds to infinitely many possible values for $w,w_a$. We compute the Jacobian, which is
\begin{align}
        J_{w, \gamma} &= 1/0.05,
        \\
        J_{w_a, \gamma} &= 2/0.05,
        \\  
        J_{p,p'} &= \delta_{p,p'} && \text{for all other } p,p',
\end{align}
with $\delta_{p,p'}$ the Kronecker delta. The new Fisher matrix with information on $\gamma$ instead of $w,w_a$ is then
\begin{equation}
        F' = J^\tnp \, F J.
\end{equation}
We can then invert the Fisher matrix in new coordinates to obtain constraints on $\gamma$. Note that while the fitting formula used is accurate for cosmologies with $w$ a function of time, it is not valid for general modified GR cosmologies. Thus these constraints consider only variations of cosmology within the framework of evolving dark energy cosmologies with standard gravity.
\end{document}